\documentclass{article}

\usepackage{microtype}
\usepackage{graphicx}
\usepackage{subfigure}
\usepackage{booktabs} 

\usepackage{hyperref}

\usepackage[accepted]{mlsys2022}

\mlsystitlerunning{Supporting Massive DLRM Inference through Software Defined Memory}

\newcommand{\fm}{\textit{FM}~}
\newcommand{\sm}{\textit{SM}~}
\newcommand{\igreel}{\textit{M1}~}
\newcommand{\ctrinsta}{\textit{M2}~}
\newcommand{\adsother}{\textit{M3}~}
\newcommand{\Tsix}{\textit{HW-L}~}
\newcommand{\Tone}{\textit{HW-S}~}
\newcommand{\Tthree}{\textit{HW-SS}~}
\newcommand{\Tseventeen}{\textit{HW-AN}~}
\newcommand{\Tseventeenoptane}{\textit{HW-AO}~}
\newcommand{\Tsixteen}{\textit{HW-FA}~}
\newcommand{\Tsixteenoptane}{\textit{HW-FAO}~}

\begin{document}

\twocolumn[
\mlsystitle{Supporting Massive DLRM Inference through Software Defined Memory}



\mlsyssetsymbol{equal}{*}

\begin{mlsysauthorlist}
\mlsysauthor{Ehsan K. Ardestani}{Facebook}
\mlsysauthor{Changkyu Kim}{Facebook}
\mlsysauthor{Seung Jae Lee}{Facebook}
\mlsysauthor{Luoshang Pan}{Facebook}
\mlsysauthor{Valmiki Rampersad}{Facebook}
\mlsysauthor{Jens Axboe}{Facebook}
\mlsysauthor{Banit Agrawal}{Facebook}
\mlsysauthor{Fuxun Yu}{gmu}
\mlsysauthor{Ansha Yu}{Facebook}
\mlsysauthor{Trung Le}{uic}
\mlsysauthor{Hector Yuen}{Facebook}
\mlsysauthor{Dheevatsa Mudigere}{Facebook}
\mlsysauthor{Shishir Juluri}{Facebook}
\mlsysauthor{Akshat Nanda}{Facebook}
\mlsysauthor{Manoj Wodekar}{Facebook}
\mlsysauthor{Krishnakumar Nair}{Facebook}
\mlsysauthor{Maxim Naumov}{Facebook}
\mlsysauthor{Chris Peterson}{Facebook}
\mlsysauthor{Mikhail Smelianskiy}{Facebook}
\mlsysauthor{Vijay Rao}{Facebook}

\end{mlsysauthorlist}

\mlsysaffiliation{Facebook}{Facebook, Menlo Park, USA}
\mlsysaffiliation{gmu}{George Mason University}
\mlsysaffiliation{uic}{University of Illinois Chicago}

\mlsyscorrespondingauthor{Ehsan K. Ardestani}{ehsanardestani@fb.com}

\mlsyskeywords{Machine Learning, MLSys, DLRM, NVM, Tiered Memory, Software Defined Memory}

\vskip 0.3in

\begin{abstract}
Deep Learning Recommendation Models (DLRM) are widespread, account for a considerable data center footprint, and grow by more than 1.5x per year. With model size soon to be in terabytes range, leveraging Storage Class Memory (SCM) for inference enables lower power consumption and cost. This paper evaluates the major challenges in extending the memory hierarchy to SCM for DLRM, and presents different techniques to improve performance through a Software Defined Memory. We show how underlying technologies such as Nand Flash and 3DXP differentiate, and relate to  \textit{real world} scenarios, enabling from 5\% to 29\% power savings. 
\end{abstract}
]



\printAffiliationsAndNotice{}  

\section{Introduction}

Recommendation models are ubiquitous across web companies \cite{netflix_reco, covington2016deep, goog_widedeep, amazon_reco, yifei2020, lopez2021bandits}, with ranking and click through rate (CTR) prediction~\cite{hazelwood2018applied, park2018deep, udit9065589} being among the widely deployed use cases. Such use cases account for a considerable demand in infrastructure resource~\cite{zhao2019recommending, gupta2020architectural, naumov2020deep} and rapid increase in data-centers footprint~\cite{anderson2021first}.

Deep learning recommendation models (DLRMs) are often composed of sets of \textit{fully connected layers (MLPs)} and \textit{embedding tables}~\cite{DLRM19}, and tend to be very large with up to trillions of parameters. One of the main reasons for such high number of parameters is that more sparse features (materialized through embedding tables) usually result in better model quality~\cite{park2018deep}. Hence the model size is mainly dictated by the \textit{embedding tables}, which could account for 100s of Gigabytes at the time of serving (inference), and 
increases rapidly year over year (e.g. 1.5x per year~\cite{jouppi2021ten} or more).

\begin{figure}[ht]
\label{fig:size_bw_dist}
\begin{subfigure}
\centering
\includegraphics[scale=0.3, angle=-90]{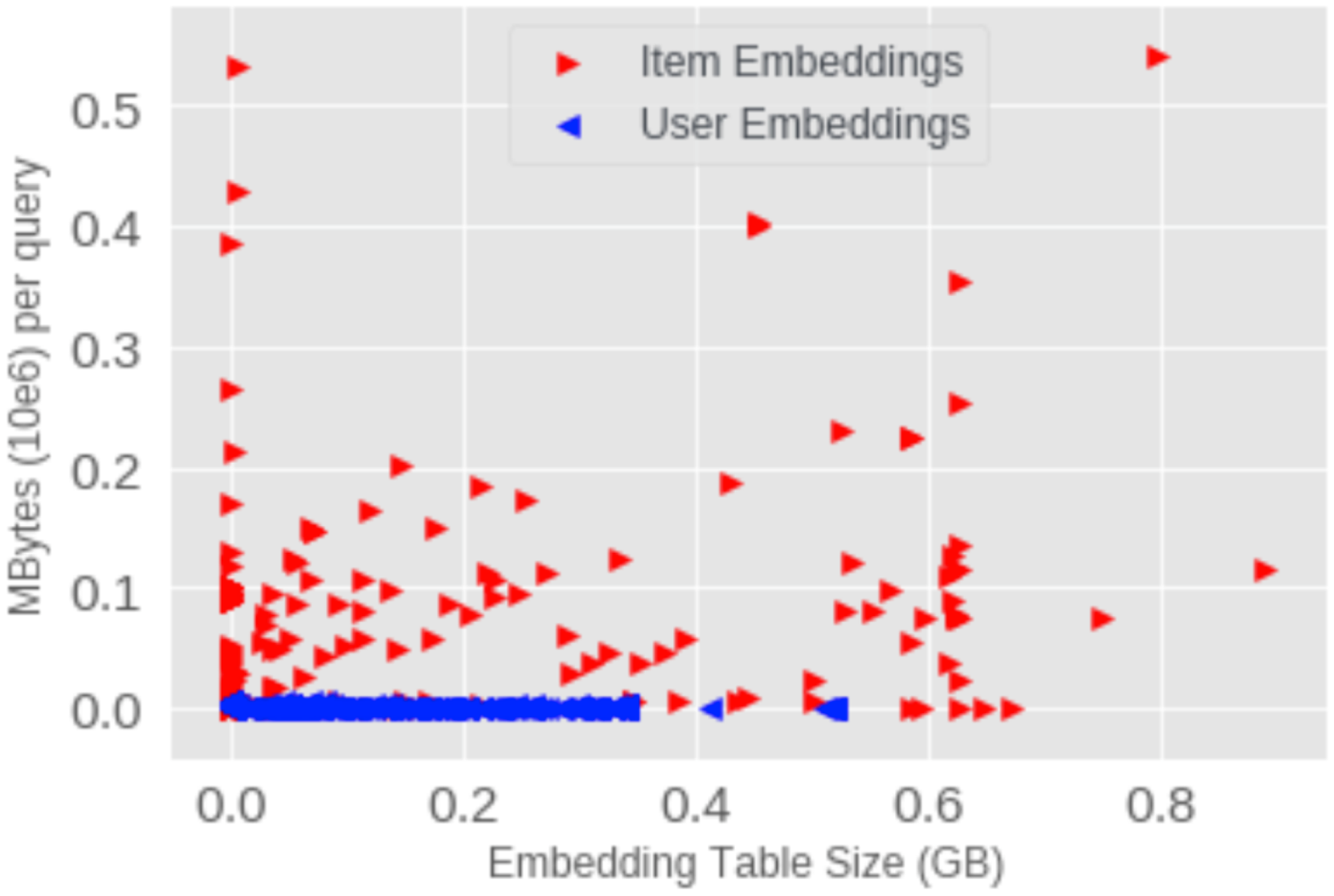}
\end{subfigure}
\caption{\textbf{Embedding Table Size (x-axis) and Bytes per query (y-axis) in a 140GB model. The model has 734 tables, out of which 445 are user tables accounting for 100GB. Majority of tables, and hence model capacity, requires low BW.}}
    \vspace{-3mm}
\end{figure}


The massive size of DLRM models requires considerable amount of memory capacity to serve. Relying on DRAM is expensive. Interestingly, not all such capacity is required at the same memory bandwidth (BW). There is a high variation in BW and Size among the embedding tables with some being accessed many times per query (e.g. 1000 accesses per query, hence requiring high memory BW), while others do not (e.g. 1 access to a row hence low BW requirement). The inherent difference between batched accessing in \textit{user} related embedding tables and \textit{item} related embedding tables (explained in Section~\ref{sec:background}) further skews such BW requirement, resulting in majority of capacity to require much smaller BW compared to a subset (mainly \textit{item} related ones) requiring high BW. Figure~\ref{fig:size_bw_dist} shows an example of such skew.

Presence of locality accessing the embedding tables (Section~\ref{sec:locality}) would further allow for leveraging slower, but denser memory through caching~\cite{eisenman2018bandana}.


Extending the memory hierarchy beyond DRAM to include slower memory technologies, such as Storage Class Memory (SCM), provides a cheaper and more power efficient approach to increase memory capacity per host. Considering the scale of deployment, the power saving could be in the order of 10s of Mega Watts. Given the importance of power in serving such models, it is becoming increasingly appealing to leverage a tiered memory. However, given the latency and BW requirement, and access granularity issues, deploying such a solution is challenging. 

This paper presents a software defined memory system which extends the memory hierarchy to SCM to accommodate the ever increasing memory capacity needs of massive DLRM models at inference. To our knowledge, this is the first paper that not only entertains the possible solutions to some aspects of enabling such technology for inference, but also evaluates all the challenges that need to be addressed by pushing the solution all the way to \textit{real world} datacenter deployment, and evaluating how the solution adds value for the end to end warehouse scale usecase.

The contributions of this paper are as follows:
\begin{itemize}

\item Extends memory available to DLRM using SCM through a Software Defined Memory stack, which can leverage different underlying technologies such as Nand Flash and Optane SSD.

\item Enables smaller granularity of read access, down to dword, for NVMe devices, which saves latency and BW, and avoids read amplification.

\item Evaluates pooled embedding cache to improve performance by bypassing dequantization and pooling when possible, and considers a range of trade offs with the cheaper capacity in slower memory to gain further performance when possible, namely de-quantization and de-pruning at load time.

\item Presents end to end results for running the usecases, and discusses the added value of the solution in realistic warehouse scale deployment scenarios.

\end{itemize}

\section{Background}
\label{sec:background}

\subsection{DLRM Architecture}
\label{sec:dlrm_arch}
Recommendation models rank a set of items according to a user's preference. For example, Amazon uses the recommendation models for selecting items in its catalog \cite{amazon_reco, yifei2020, lopez2021bandits}, Netflix for showing movie options \cite{netflix_reco}, Google for displaying personalized advertisements \cite{goog_widedeep}, and Facebook for ranking and click through rate (CTR) prediction \cite{hazelwood2018applied, naumov2020deep}. 

Deep Learning Recommendation Models~\cite{DLRM19} are often composed of two main components:

\begin{enumerate}
    \item Embeddings, which map the categorical features (e.g. what subject a user has shown interest in) into dense representations. Different categorical features have varying cardinality, and hence require different size when materialized through embedding tables. The embeddings could be further divided into \textbf{user embeddings} (materializing categorical features for users) and \textbf{item embeddings} (materializing categorical features for items to be recommended such as news and movies). The embeddings are typically memory intensive. 
    \item Interaction, which aggregates continuous features and the dense representation of categorical features (e.g. by concatenation), and captures their complex interaction (e.g. by multi-layer perceptrons (MLP)). The interaction components are typically compute intensive.  
\end{enumerate}

\begin{figure}[t]
\centering
\includegraphics[scale=0.3, angle=-90]{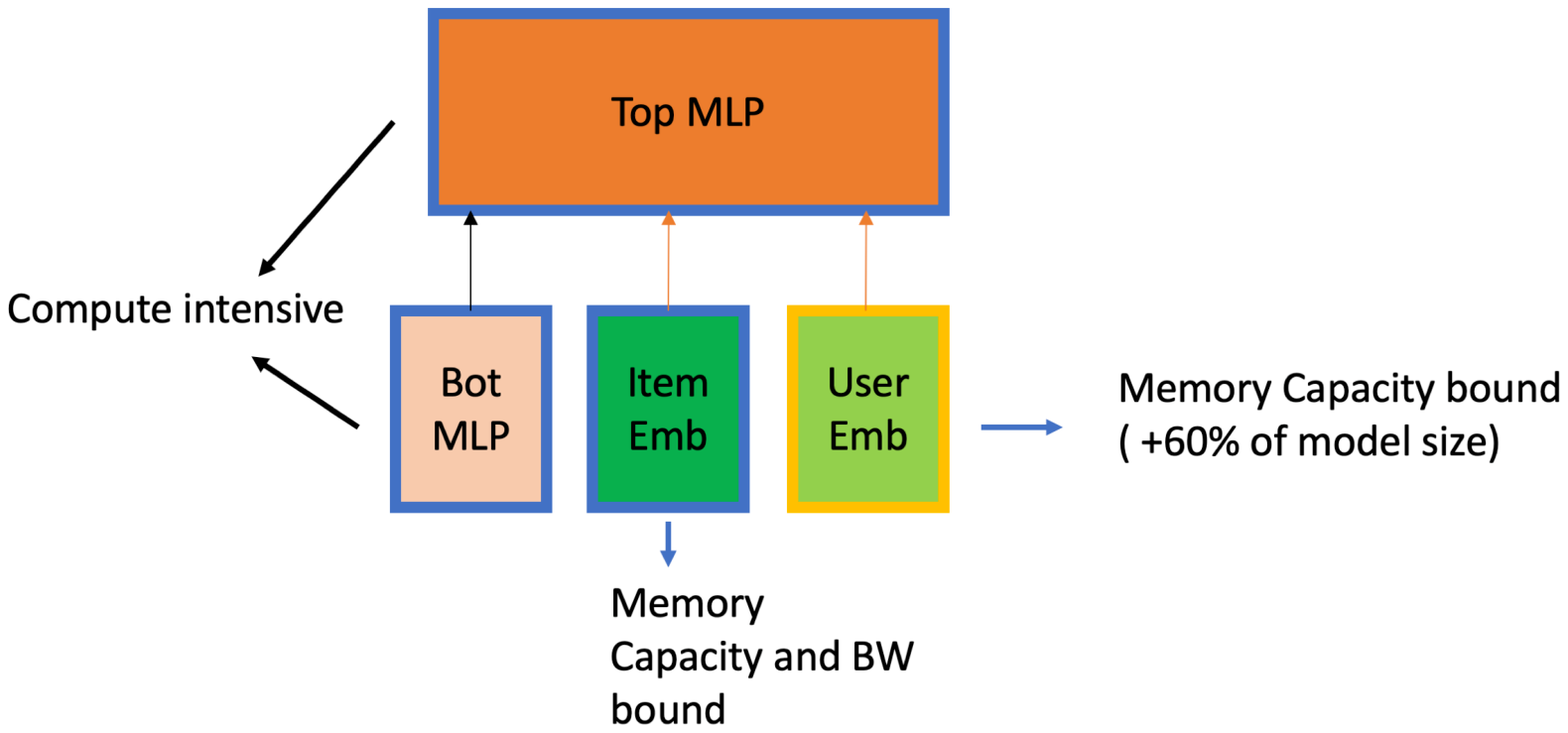}
\caption{\textbf{High level DLRM architecture.}}
\label{fig:dlrm_arch}
\end{figure}

Figure~\ref{fig:dlrm_arch} depicts the high-level architecture of DLRM models. Bottom MLP reprojects the continuous features (e.g. age of the user) to dense ones. The embeddings components convert the categorical features to dense representation, and the top MLP captures the interaction of all the features.

\subsection{BW and Capacity Requirement}
\label{sec:bw_cap}

BW requirement for embeddings can be defined as 
\begin{equation}
\label{eq:bw}
\footnotesize
 BW = QPS * \sum (p_i * d_i), i \epsilon T
\end{equation}

where \textbf{QPS} is Query Per Second, which is the rate at which the inference queries are expected to be processed in a given host, $p_i$ is the \textbf{pooling factor} (number of embedding rows which needs to be looked up per query) for table $i$, and $d_i$ is the \textbf{embedding dimension} for table $i$. $T$ is the number of embedding tables in the model. Note that the number of rows in the tables does not impact BW.

For inference, several items will be evaluated (ranked) to arrive at the top items for recommendation. Hence an inference query could access the user embeddings once for that user, while accessing the items' embeddings for a batch of items. The user side embedding results could be broadcasted to all the items for Top MLP computation. It's worth nothing that this is different from training where one input sample consists of one user and one item. As a result, the \textit{BW requirement per query for user embeddings is much lower than that of item embeddings}. We can rewrite Equation~\ref{eq:bw} as follows:  

\footnotesize
\begin{equation}
\label{eq:bw_ro}
 BW = QPS * ( B_I\sum (p_i * d_i) + B_U\sum (p_j * d_j) ),\\
  i ~\epsilon~ T_I,  ~~ j ~\epsilon~ T_U
\end{equation}
\normalsize

where the batch size for Items and Users ($B_I$ and $B_U$, respectively) is separated. $T_I$ and $T_U$ denote the number of item and user embedding tables, respectively. Given the latency sensitivity of inference queries, $B_U$ is typically 1. $B_I$ could in in order of 10s or 100s of items (depending on how fast they can be processed).  

Our observation shows that more than 2/3 of the model capacity are contributed by the user embeddings. This could be due to the fact that there is a wider set of categorical features to describe the users, resulting in more user embeddings being used in the model. The implication is that \textit{the bigger portion of model size have lower BW requirement}.

Another important observation is that the execution of user and item embeddings are independent, while the Top MLP has dependency on both. Assuming the user embeddings are the prime candidate to be accommodated by slower memory~\footnote{Large item tables with low pooling factor could be considered for placement on the slower memory. However, without lack of generality, in this work we primarily focus on placing user embeddings on slower memory.}, \textit{as long as the access time for user embedding is still smaller than that of item embeddings, the slower access due to slower memory is not exposed} in end to end latency. Equation~\ref{eq:lat_budget1} formulates, at high level, the time budget for the slower memory.
\begin{equation}
    \label{eq:lat_budget1}
    \footnotesize
     time(UserEmbeddings) = time(ObjectEmbeddings).
\end{equation}
which could be elaborated further as follows:
\begin{equation}
    \label{eq:lat_budget2}
    \footnotesize
     BW_q(user)/BW_{SlowMem} = BW_q(items)/BW_{FastMem} 
\end{equation}

$BW_q(user)$ refers to the BW requirement at a given query for user embeddings, and $BW_q(items)$ denotes that of item embeddings.

\subsection{Hyper-Scale Deployment}
\label{sec:hsdeployment}


\textbf{Latency and Throughput}: The inference of DLRMs are both latency and throughput sensitive. The latency sensitivity is derived from real time user interaction, requiring the latency in 10s of millisecond range for the ranking. At the same time, queries at Data Center level need to be processed within the expected throughput. Given QPS per host at a given target latency, the total throughput (e.g. in a DC region) will translate into a set number of hosts (Equation~\ref{eq:qps}-\ref{eq:num_host}). Note that the latency requirement varies across different model/usecases. For example some models have strict p99\footnote{Here, p99 denotes 99 percents of queries needs to be processed within the latency requirement, similar for p95.} latency requirement with active load balancing to ensure the latency requirement across the fleet. Other models/usecases could have desired p95 latency which is achieved through static allocation of resources.

\begin{equation}
\small
    \label{eq:qps}
    QPS(HW) \propto\\
    min(BW(HW)/BW_q, ~Comp(HW)/Comp_q)
\end{equation}
\begin{equation}
\small
    \label{eq:lat}
    Latency(HW) \propto \\
    sum(BW_q/BW(HW), ~Comp_q/Comp(HW))
\end{equation}
\begin{equation}
\small
    \label{eq:num_host}
    Resources(HW) \propto QPS_{Total} / QPS(HW)~~~~~~
\end{equation}
\normalsize

\textbf{Scale Up vs Scale Out}: As the model size increases, either the memory per host needs to increase (scale up) or the model needs to be sharded to scale the memory by leveraging multiple hosts (scale out, e.g. see ~\cite{lui2021understanding}). Extending the memory to SCM could be considered as a scale up only approach, or applied to the hosts involved in scale out to reduce the fan out. 
It needs to be mentioned that increase in model size usually is accompanied with increase in compute intensity of the model as well. So the relevant approach to serve the model would depend on the compute, memory BW and memory capacity requirement and their relative ratio. 

\textbf{Power Boundness}: DLRM models keep increasing in their complexity and size faster than the rate new data centers could be developed (e.g. see ~\cite{anderson2021first}). Furthermore, DLRM applications account for a considerable portion of infrastructure resources~\cite{gupta2020architectural, zhao2019recommending}. This leads to power boundness of the usecase, with query/watt at the acceptable latency being the primary metric to solve for at scale. Tiered memory directly helps with this top line metric by 1) leveraging more power efficient per GB memory when possible, and 2) allowing for better system solution, e.g. not scaling out. 

 \textbf{TCO}: Another important factor impacted by the choice of tiered memory is the Total Cost of Ownership (TCO). Cheaper memory per GB reduces the total TCO. Furthermore, higher memory capacity per host (scale up) is not always possible~\footnote{For example given limited number of DIMM slots per CPU (e.g. 6), with a maximum capacity per DIMM (e.g. 128GB), there is a max DRAM amount (e.g. 768GB) that could be deployed.}, which could require tiered memory or scaling out solution. Such choices shape the end to end \textit{system} serving the model, with different overhead and efficiency, and consecutively impact the power provisioning and TCO beyond component level.

\section{Technology}
\label{sec:scm_choice}
Extending the memory hierarchy to SCM can be deployed regardless of the choice of accelerators (e.g. using GPU for inference), and the hierarchy of faster memories (e.g HBM + DRAM). Hence, we refer to the the last level memory with SCM as \sm (for slow memory) and the first level(s) of memory as \fm (for fast memory). 

\begin{figure}[t]
\centering
\includegraphics[width=\linewidth]{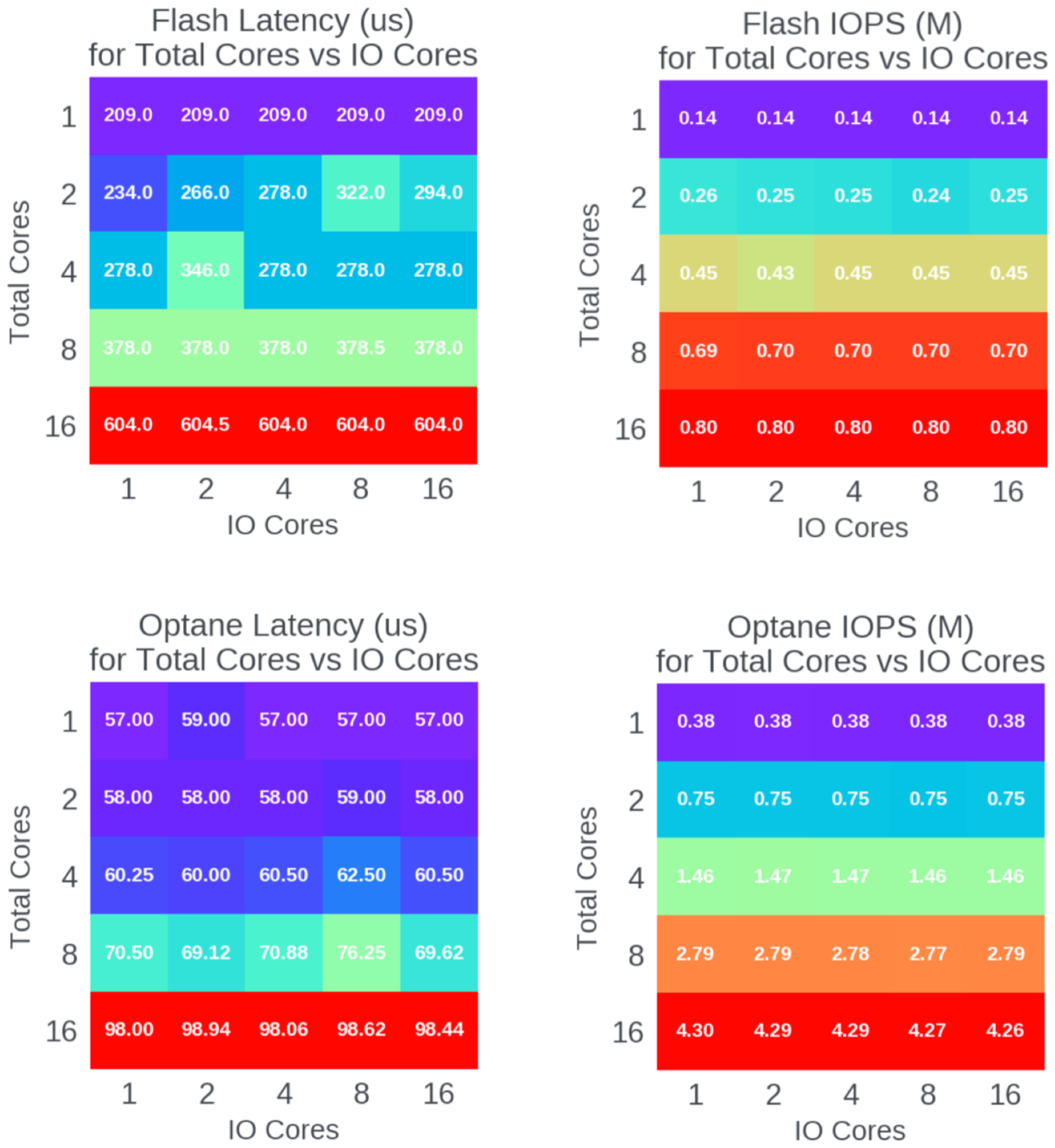}
\caption{\textbf{IOPS and latency for Nand Flash and Optane SSDs. Given each query to a table involves multiple lookups (pooling factor), we benchmark each device with average of 20 lookups per IO. The latency is for the batch of 20 lookups. As the results show, Optane SSD provides much lower latency and higher IOPS than the Nand Flash.}}
\label{fig:scm_perf}
\end{figure}

\begin{table*}
\caption{\small{Different options for the slow memory (\sm). The values are based on public information. Cost is relative to DRAM}}
\vspace{2mm}
\footnotesize
\label{tab:scm}
\centering
\setlength{\tabcolsep}{4.2mm}{
\begin{tabular}{l|c|c|c|c|c|c}
\toprule
Technology & IOPS (M) & Latency (us) & Endurance & Access Granularity & Cost & Sourcing \\ \midrule
PCIe Nand Flash & 0.5 & O(100) & ~5 & 4K & 1/30 & multi   \\ \midrule
PCIe 3DXP (Optane) & 4 & O(10) & 100 & 512 & 1/5 & single  \\ \midrule
PCIe ZSSD & 1 & O(100) & ~5 & 4K & 1/10 & single \\ \midrule
DIMM 3DXP (Optane) & - & O(0.1) & - & 64 & 1/3 & single  \\ \midrule
CXL 3DXP & $>$10 & O(0.5) & - & 64-128 & - & single \\ \midrule
\end{tabular}}
\end{table*}

There is a range of technologies that could be used for \sm. Table~\ref{tab:scm} lists some of the currently more readily options. We track the following key parameters for each technology:

\textbf{IO Per Second (IOPS)}. The access patterns to the embedding tables are random. We track IOPS instead of GB/s, because the embedding rows, and hence the access granularity to the \sm, is typically much smaller than 4KB block size (read amplification). The inference access is read only, with non-frequent writes only during model update.

\textbf{Access Granularity}. The quantized embedding rows, while growing, are in 128-256B range~\footnote{We primarily use row-wise quantization}. IO Read with higher granularity (e.g. 4KB) will result in read amplification and wasted BW.

\textbf{Latency}. This is the loaded access latency for a block of data. Different technologies show different curve as the load increase from low to high. Given latency sensitivity of the usecase, we need to operate on a latency region that is in order of up to a few 10s of us. 

\textbf{Write BW}. The only write access happens during model update. In general more symmetric read and write BW becomes more important as the update frequency increase.

\textbf{Endurance}. The endurance can translate to model update interval. ($Update Interval = 365 * ModelSize / (pDWPD~\footnote{Physical Drive Write per Day} * \sm Capacity)$

\textbf{Cost}. Relative cost per GB compared to DDR4 DRAM

\textbf{Sourcing}. How many vendors offer the technology. The higher, the better.

Nand Flash provides the cheapest option, with multiple vendors offering the technology. However, it suffers from two drawbacks. Low random IO per second (IOPS), and increased latency as the IOPS increases (Lower endurance can be offset by capacity). This limits the usecase to models with low BW requirement. 

PCI3 3DXP (Optane) provides a good random IOPS (4M at 512B) and considerably better latency profile compared to Nand flash (O(10) usec). The endurance is also high enough to accommodate frequent updates. As a result, Optane SSD can enable tiered memory for the frontier of the models with high capacity and BW requirements. Figure~\ref{fig:scm_perf} shows the IOPS and latency profile for Nand Flash and Optane SSD.

PCIe ZSSD offers better latency compared to Nand Flash, but does not offer high enough IOPS to set it considerably apart from Nand Flash.
DIMM 3DXP impacts the available memory BW to the CPU which is point of concern. CXL 3DXP would provide the best performance in the set, without having the negative side effect of DIMM 3DXP. But still not as readily available as other technologies listed here. 

The choice of technology for \sm depends on specific usecase and model characteristics. As the models scale size and BW, the higher BW options become more relevant. We observe that Nand Flash and Optane SSD enable tiered memory for a wide range of DLRM models (lower-end with less strict p99 latency, and higher-end of BW requirement, respectively). As the model's capacity and BW scale overtime, CXL based solution would become more relevant.  

\section{Design and Implementation}
\label{sec:design}
We evaluated several different design choices for the software stack. Given the scheme could be used for a wide range of model configuration and underlying HW, we evaluate the design choices, such as cache organization, by evaluating a wide range of target models beyond what presented in the results section. We also consider both Inference as well as Inference Eval (see Table~\ref{tab:usecase}). This is to avoid over designing for a particular usecase. Several tuning options are provided such that the desired serving configuration could be decided at model deployment time (e.g. through an auto-tuning tool). Such tuning options are highlighted as \textit{Tuning API} in each subsection.

\begin{table}[]
\footnotesize
\caption{\small{Usecases}}
\vspace{2mm}
\label{tab:usecase}
\setlength{\tabcolsep}{3.5mm}{
\begin{tabular}{l|l}
\toprule
Inference & user batch size = 1, \\
& item batch size $>$ 1 (O(100)), \\
& Inference is latency sensitive. \\ \midrule
InferenceEval* & The goal is accuracy validation.\\
& user batch size == item  batch size $>$ 1. \\ 
\bottomrule
\end{tabular}}
\vspace{0.5mm}
\\
*InferenceEval is similar to eval after training, but model has gone through inference specific transformation. 
\end{table}

\subsection{Fast IO}
\label{sec:fast_io}
Most of the relevant \sm technologies currently are block devices with NVMe interface. As the BW requirements of the model grow, the IOPS requirement consequently grow (Equation~\ref{eq:bw_iops}). However, IO through NVMe stack is still an expensive operation. Performing multi-million IO per second could required prohibitive amount of computing resource (CPU). We have chosen to use ~\textit{io-uring}~\cite{iouring} due to its lower overhead per IO. Figure~\ref{fig:scm_perf} shows the performance characteristics of PCIe Nand Flash and PCIe Optane using io-uring.

\begin{equation}
\label{eq:bw_iops}
\footnotesize
    IOPS \propto QPS * \sum (pi),~~ i ~\epsilon~ Tables(\sm)
\end{equation}

One particular design choice was $mmap$ vs $DIRECT-IO$. Due small access granularity and lack of considerable spacial locality (Section~\ref{sec:locality}), we observed that $mmap$ would not provide the best use of \fm space, and results in higher access latency (by 3x. e.g. reading in and maintaining 4KB into memory for a 128B request). Hence we opted for $DIRECT-IO$ with an application level cache.

Given different technologies could be used for \sm, we realized some optimizations are technology specific. For example with Nand Flash, we need to smooth out the bursts by limiting the maximum outstanding requests to the SSD because SSD controllers typically try to serve all possible outstanding requests which results in extra latency.

\textit{Tuning API}: Total number of outstanding IOs per table and total number of tables that can be processed at given time.

\subsubsection{Enabling small access granularity}
Sub-block (e.g. 4KB) reads is not normally supported by an operating system. The higher access granularity to the SCM device, given the lack of spacial locality (Section~\ref{sec:locality}) has three adverse implications: 
1) higher latency due to read amplification as more data need to be transferred from device to the host; 
2) more pressure on the interconnect (PCIe) in the system, which might require provisioning more PCIe lanes, and hence increased system power and cost;
3) requiring extra memory copy to handle extracting row data from block data and copying it into the cache. Given that majority of tables have embedding dimension smaller than 512B at inference (due to quantization~\cite{guan2019post}), we have enabled arbitrary access granularity, down to DWORD, with NVMe. A two legged approach is taken to achieve this goal. 

\begin{itemize}
    \item Linux Kernel: Linux kernel is updated to allow a custom command over the \textit{io-uring}~\cite{iouring} application transport that allows down to 4B granularity reads.
    \item NVMe Driver: The NVMe Scatter Gather List (SGL) Bit Bucket is used to communicate the desired portion of a block. This allows full flexibility as to in which parts of a request the host is interested, hence only transferring the necessary parts of a read over the bus.
\end{itemize}

By only reading the parts of a block that is necessary, we save around 75\% of the bus bandwidth and reduce the time needed to transfer this data. This reduces the observed latency of a given read by $~$3-5\%. The savings at the application level are more given removal of the extra memcpy (see Section~\ref{sec:cache_org} for more details). 

Both of these features will be submitted for the upstream kernel, and will be publicly available.

\begin{figure}[]
\small
\centering
\begin{subfigure}[]{}
\includegraphics[width=0.6\linewidth]{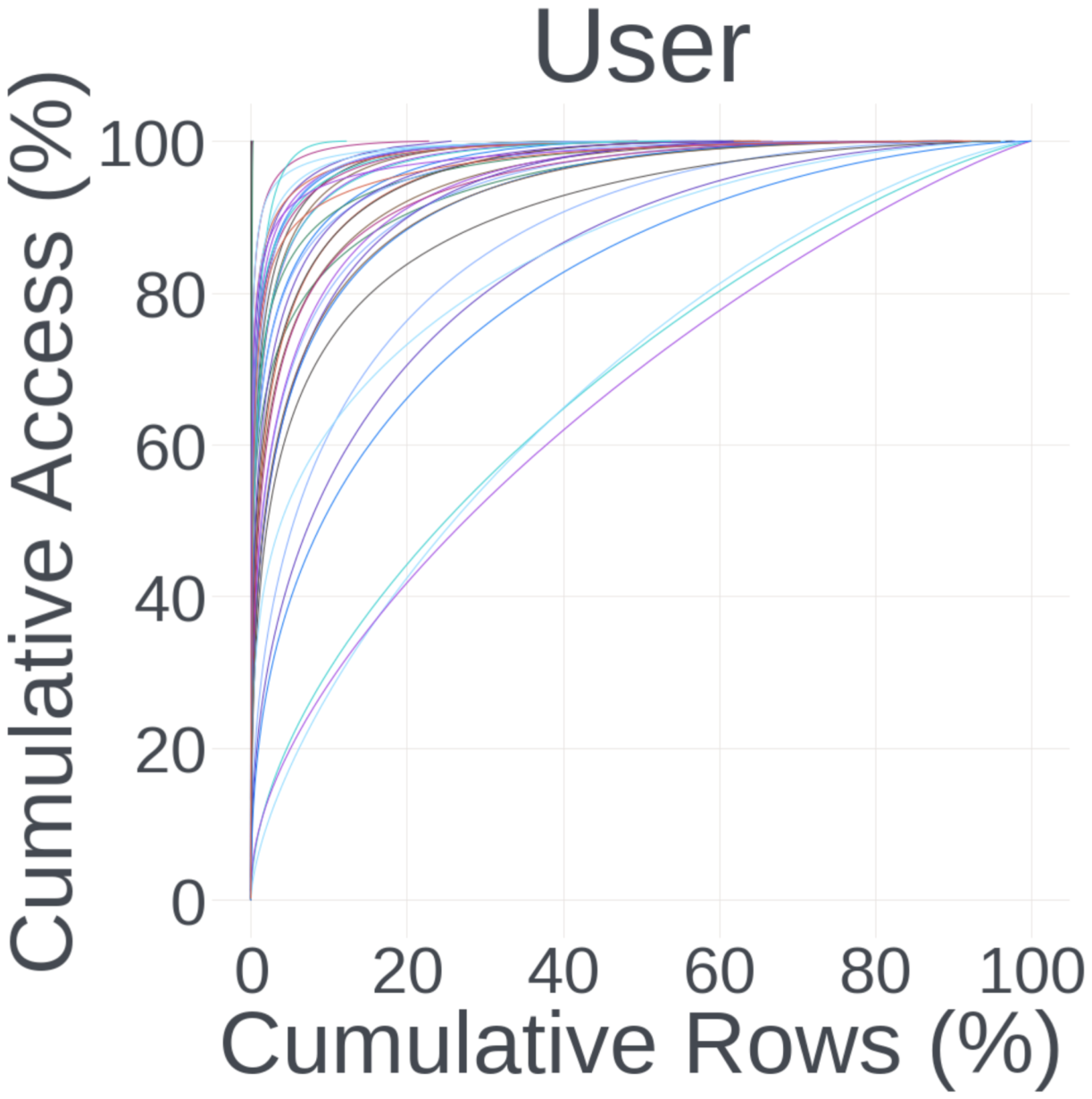}
\end{subfigure}
\begin{subfigure}[]{}
\includegraphics[width=0.6\linewidth]{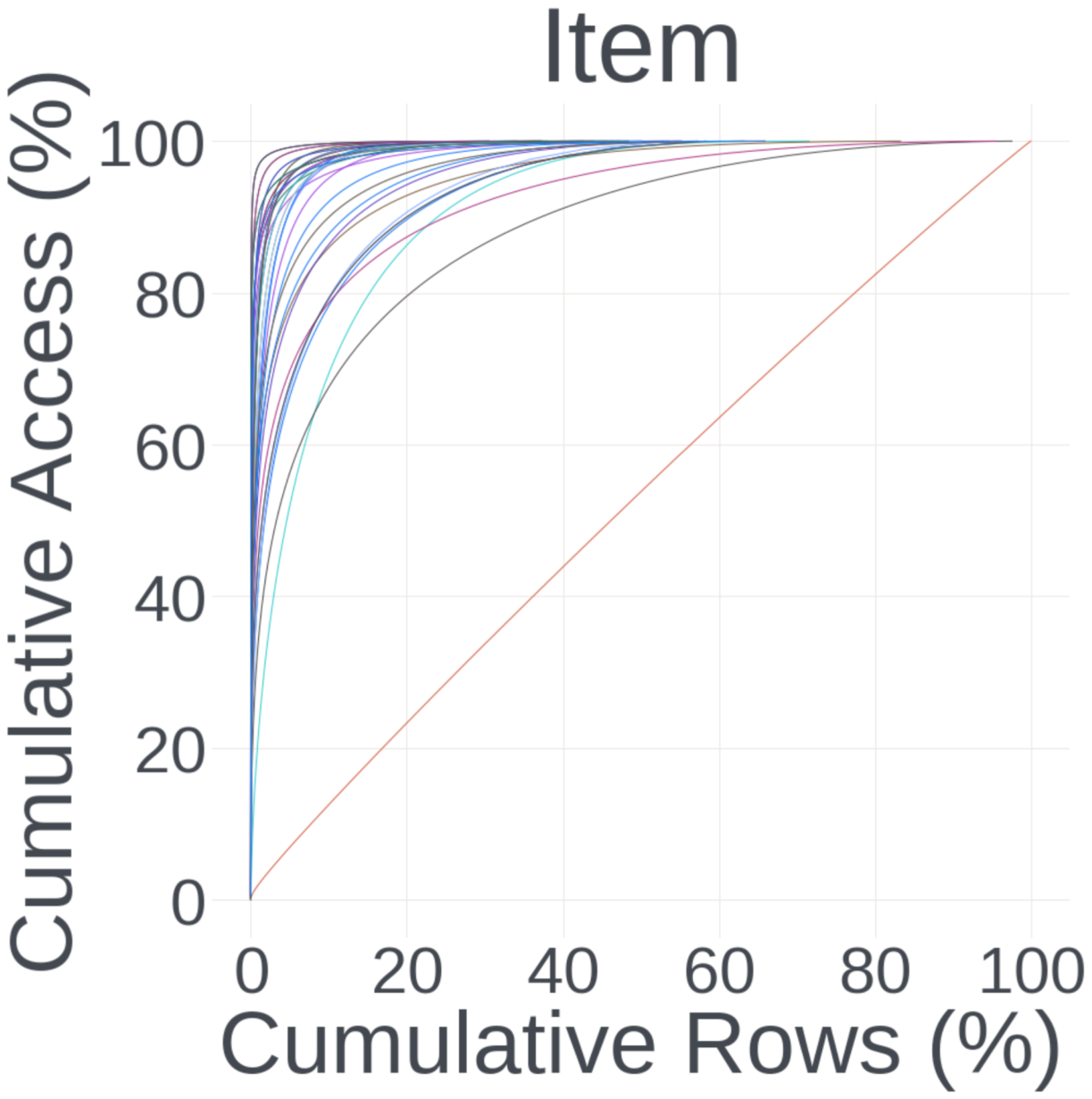}
\end{subfigure}
\begin{subfigure}[]{}
\includegraphics[width=0.6\linewidth]{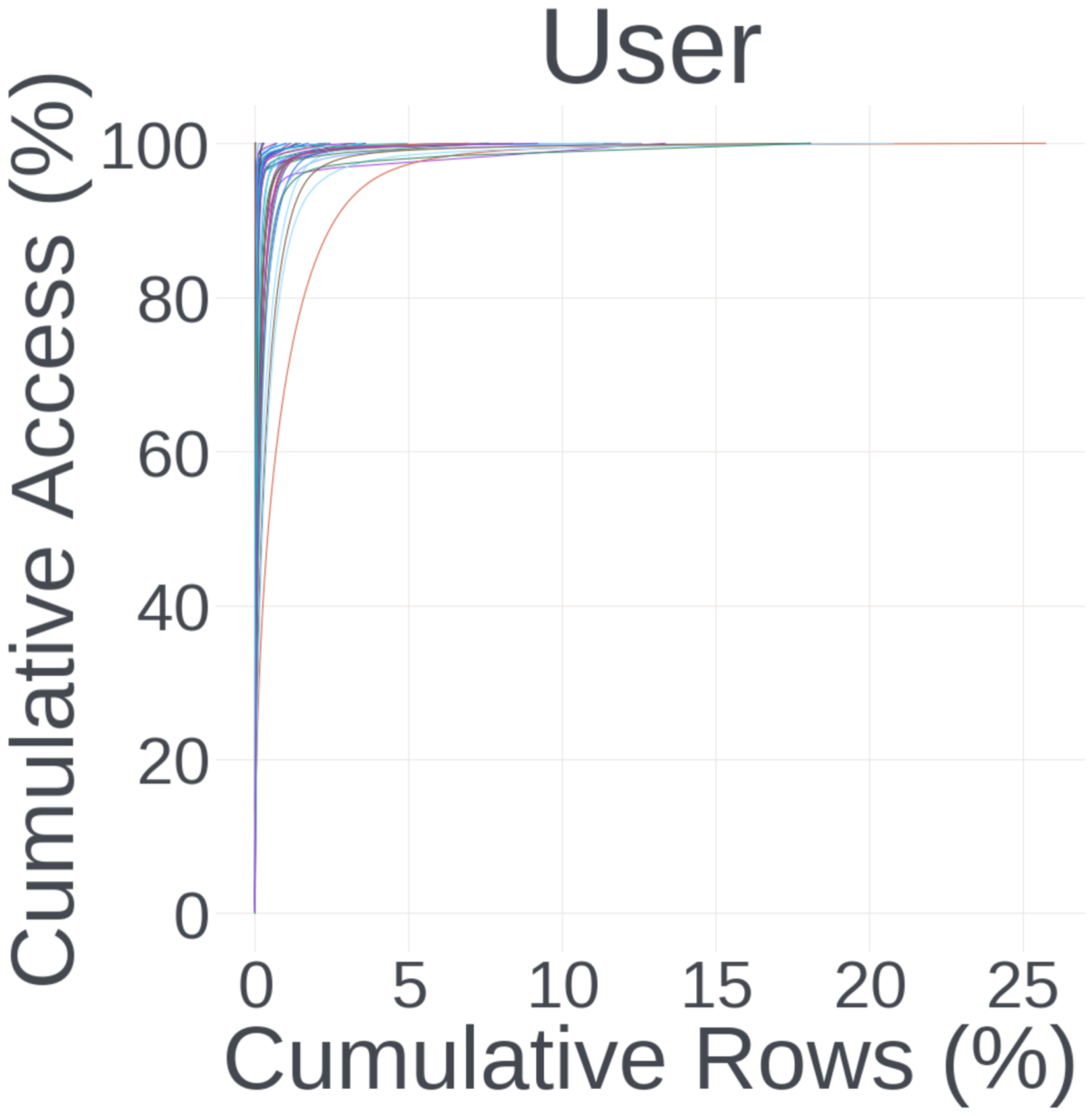}
\end{subfigure}

\caption{\textbf{Temporal Locality accessing User (a) and Item (b) embeddings. Access to majority of the tables demonstrate power law. For each plot, we track 50 tables at random, for data sampled post hash for 6 days. (c) shows temporal locality for the same set of user tables observed by one host during serving, indicating higher locality.}}
\label{fig:locality_cdf}
\end{figure}

\begin{figure}[]
\small
\includegraphics[scale=0.3, angle=0]{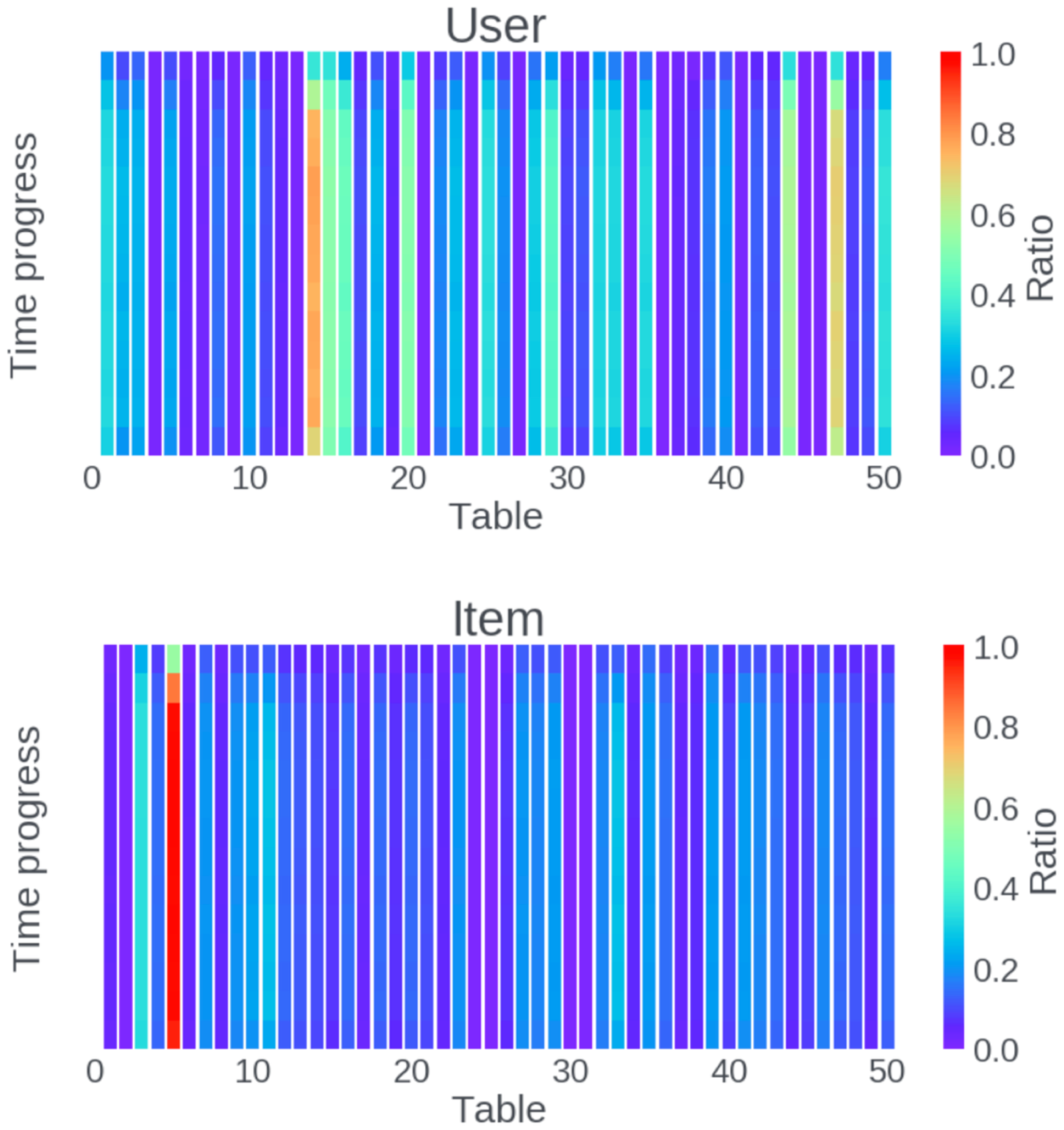}
\caption{\textbf{Spatial Locality accessing User and Item embeddings. Value 1.0 indicate 100\% spatial locality. For each plot, we track 50 tables at random, for data sampled for 6 days. The average window is around 25M access per table}}
\label{fig:locality_spat}
\end{figure}

\subsection{Locality}
\label{sec:locality}
Locality is an important characteristic of accessing embedding tables as it could allow for providing a higher effective BW for the data in \sm by a cache in \fm. 
Figure~\ref{fig:locality_cdf} captures \textit{temporal locality} through the cumulative distribution of a range of categorical features. Majority of the features show a power law distribution, with a small subset of embedding rows accounting for majority of accesses, hence high temporal locality. We separate User and Item embeddings since we observe a meaningful difference in the distributions (item embeddings show more locality). This motivates the use of a Software Managed Cache in \fm to cache the hot portion of embeddings placed in \sm.  

Note that the temporal locality observed from a host also depends on the serving system. 
Inference queries will go through a scheduler/aggregator which routes a query to a specific host for ranking. Figure~\ref{fig:locality_cdf}-(c) shows the temporal locality for the same set of user embedding tables, but observed from one host during serving, which shows higher locality. Enforcing a user-to-host sticky policy can help increase cache hit rate observed from a host.

Figure~\ref{fig:locality_spat} demonstrates the degree of \textit{spatial locality} accessing the table. It uses the average ratio of unique index to unique 4KB block size, normalized to the maximum unique index per block size per table, as proxy for spacial locality. The ratios are captured in intervals (average 25M access per table). Value 1.0 indicates the same number of unique index and unique 4KB blocks, i.e. high spacial locality. The heat map and the cooler temperature overall indicates low spatial locality.

\subsection{Cache Organization}
\label{sec:cache_org}

The design and organization of the cache also has impact on the overall performance. 

\textbf{Unified Row Cache}: Given the observation from the locality study (Section~\ref{sec:locality}), we opted for a unified row cache. The unified cache allow for better utilization of the memory space compared to per table cache, and the lack of considerable spacial locality motivated the row cache. We use CacheLib~\cite{berg2020cachelib}. Without the small granularity access (Section~\ref{sec:fast_io}) we need to copy a block of data into an aligned \fm buffer, and then copy the desired portion into the cache. This means more than 2X \fm BW needed for every X data pulled in from \sm, and increased latency due to the memory copies. With small granularity access, we can directly copy data to the cache storage, and save on \fm BW and improve latency.

\begin{figure}[t]
\centering
\small
\includegraphics[scale=0.3, angle=-90]{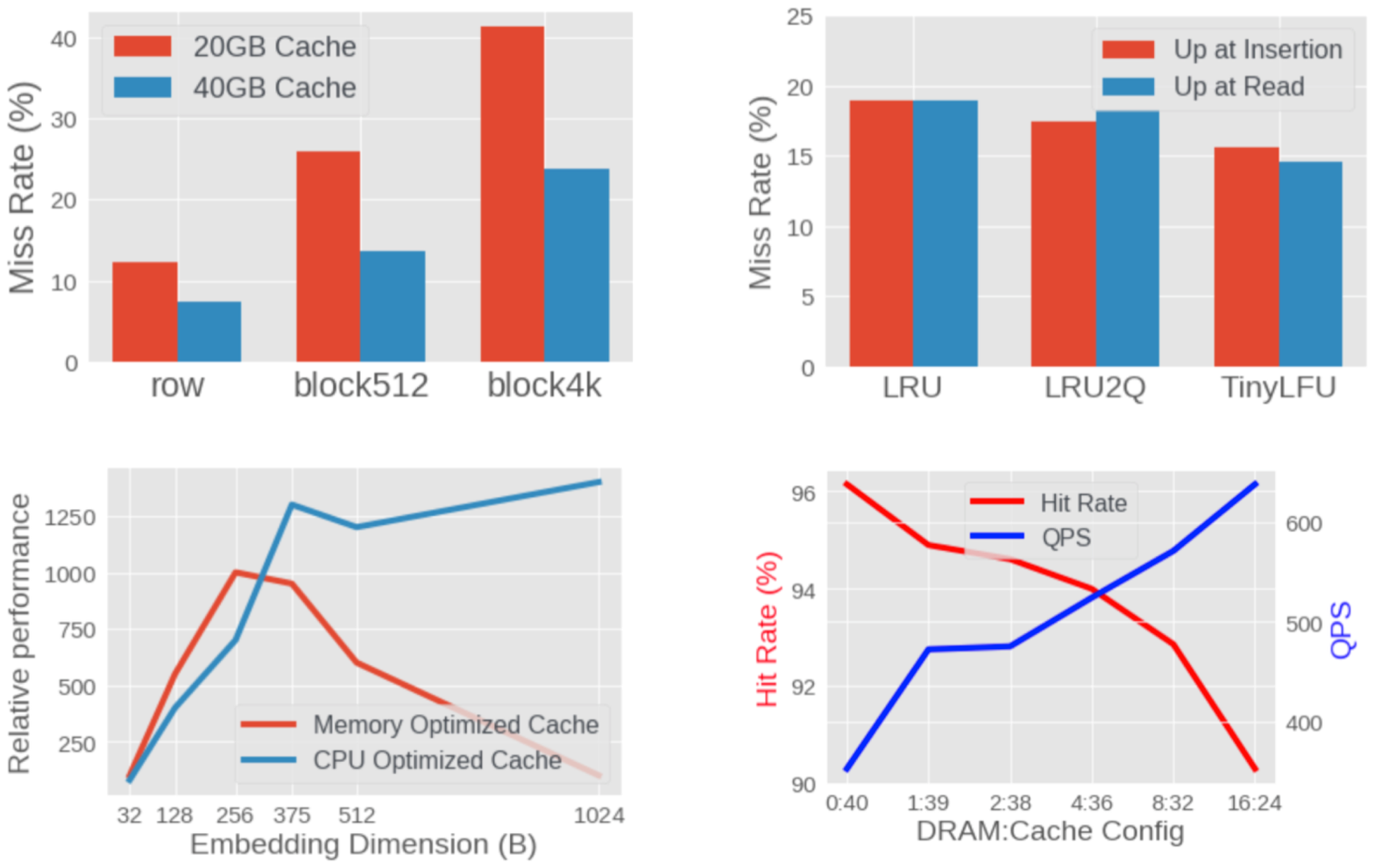}
\caption{\textbf{Performance implications of different cache organization choices. We opt for a unified row cache, which internally implements two caches optimized differently based on embedding size. The cache routes the requests to proper internal cache based on embedding dim (Embedding dim $<=$ 255 will be routed to memory optimized cache). The bottom right figure shows a case where direct placement on DRAM could have considerable impact on QPS.}}
\label{fig:cache_org}
\end{figure}

\textbf{Memory vs CPU overhead}: With CacheLib, we had the choice to tune for memory overhead (less overhead per key-value pair, but requires search in a bucket) or pay for memory overhead and optimize for CPU utilization (higher overhead per key-value pair). Majority of tables have embedding dim smaller than 256B, but there is small but a growing subset which have bigger than 256B embedding dim. Given the overhead and performance results shown in Figure~\ref{fig:cache_org}, we opted for a dual cache where tables with embedding dim smaller than 256B are routed to a Memory Optimized Cache, and to CPU Optimized Cache otherwise. 

We also evaluated multi-level cache (row cache backed by a block cache) but did not observe any benefit. 

\textit{Tuning API}: Cache sizes and number of cache partitions.

\subsection{Pooled Embedding Cache}
\label{sec:pooledemb}
\begin{table}[!t]
\caption{\footnotesize{Summary of Pooled embedding cache profiling for 100M queries. Hit rate shows the number of queries which got a hit for at least one of the subsequences. Generated sequences show the normalized number of subsequences generated. While considering $0<c<=P$ increases the cache of a hit for a subsequence, in practice the overhead of finding such subsequence is prohibitive except for $c$ near $1$ or near $P$.}}
\label{tab:pooledemb_profile}
\vspace{2mm}
\footnotesize
\centering
\setlength{\tabcolsep}{3mm}{
\begin{tabular}{l|c|c}
\toprule
Scheme & Hit rate (\%) & Generated sequences \\ \midrule
c=10 & 26 & O($avgP \choose c$) \\ \midrule
c=10, top indices & 19 & O(100) \\ \midrule
c=P & 5 & 1 \\ \midrule
\end{tabular}}
\end{table}

The \sm cache stores the raw quantized embeddings. For every embedding operator, there are $p_i$ embeddings read out for $table_i$ from the cache or from \sm, which then go through dequantization and pooling~\cite{khudia2021fbgemm} to generate the output for Top MLP. If we had the resulting pooled embeddings already cached (or even partial pooled embeddings), we could (partially) save lookup, dequantization and pooling. 

\begin{table}[!t]
    \centering
    \caption{Average length (number of indices) saved for hit in PooledEmb Cache with a 4GB cache size.}
    \vspace{2mm}
    \label{tab:pec}
    \footnotesize
    \setlength{\tabcolsep}{6.5mm}{
    \begin{tabular}{c|c|c}
    \toprule
         LenThreshold & Hit Rate & Hit Avg Len  \\ \midrule
         1 & 4.39\% & 11 \\
         4 & 4.58\% & 35 \\
         8 & 4.02\% & 40 \\
         16 & 4\% & 56 \\
         32 & 3.9\% & 76 \\
         \bottomrule
    \end{tabular}}
    \vspace{-3mm}
\end{table}

We profile queries to establish whether there is locality in sequence of indices that appear across queries. Table~\ref{tab:pooledemb_profile} shows the profiling result. There is $P \choose c$ choices for an embedding operation with $P$ indices. For a subsequence of indices of size $c, 0<c<=P$, the possibility of a repeating subsequence decreases as $c$ increases. However, except for near the edges (e.g. $c=1$ or $c=P$), the number of possible subsequences is too large. In our profiling we limit the length of subsequence to 10, and only profile most frequent indices. Nonetheless, our observation is that in the case of $c=P$ , where we only cache the full sequence and only lookup for the full sequence of indices for each table when a request arrives, provides small enough overhead, and reasonably high enough hit rate to have a chance at improving performance 
. Algorithm~\ref{alg:pooledemb_how} depicts the implementation. We observed around 5\% hit rate for the pooled embeddings (Table~\ref{tab:pec}). The average length of requests hit in PooledEmbedding Cache increase as the $LenThreshold$ is increased.

\begin{algorithm}
   \caption{Pooled Embedding Cache}
   \label{alg:pooledemb_how}
\begin{algorithmic}
    \footnotesize
   \STATE {\bfseries Input:} Table, Indices
   \STATE doPooledEmbCache = len(indices) $>$ LenThreshold
   \IF{doPooledEmbCache}
   \STATE $sequenceKey = hash(indices)$
   \IF{$e = lookup(t, sequencekey)$}
   \STATE  return e  //pooled emb vector exists in the cache
   \ENDIF
   \ENDIF
   \FOR{i in Indices}
   \IF{not E[i] = lookup(t, i)}
   \STATE $prepareIO(t, i)$
   \ENDIF
   \ENDFOR
   \STATE $submitIOs(E)$
   \STATE // dequantize and pool all the embedding vectors in the sequence
   \FOR{e in E}
   \STATE output += dequant(e) 
   \ENDFOR
   \IF{doPooledEmbCache}
   \STATE cache[sequenceKey] = output
   \ENDIF
   \STATE return output
\end{algorithmic}
\end{algorithm}

Algorithm~\ref{alg:pooledemb_how} shows the implementation. We use an order-invariant hash to create a key from the sequence of indices in a request. 

\textit{Tuning API}: The min sequence length which could be cached is configurable ($LenThreshold$). 

\subsection{\sm vs \fm capacity Tradeoff}

Given the cheaper \sm capacity, we evaluated a few approaches that reverse the schemes commonly used to reduce model size, namely \textit{de-pruniung} explained here, and de-quantization in Appendix~\ref{sec:discussion}.

Pruning embedding tables post training is commonly used to reduce inference model size (e.g. see ~\cite{lui2021understanding}). At high level, the embeddings rows with values very close to 0 are heuristically removed. A new tensor is defined to map the indices in the un-pruned space to indices in pruned space. The size of a mapping tensor is $NumRow(Unpruned) * IdxType$,  $IdxType \epsilon \{4, 8\}Bytes$. To place pruned embedding tables on \sm, we can either 1) save both the pruned table and mapping tensor to \sm, which means two accesses to \sm per embedding lookup; or 2) place the pruned table on \sm and keep the mapping tensor in \fm. Given the IOPS boundness with \sm, and relatively smaller size of the mapping tensor, options 2 is a more desirable choice. However, as the model size increases, the aggregate size of the mapping tensors increases. The space taken by mapper tensors are the memory that is taken away from the \sm cache. 

To free up the memory used by mapping tensors, we can \textit{de-prune} the embeddings at the time of loading. Algorithm~\ref{alg:deprune_code} shows how de-pruning is done. Beside increased model footprint on \sm, de-pruning could lead to extra accesses to \sm, and consequently cache pollution. This is because the pruned embeddings now will be accessed and cached. However, the intuition is that the pruned embeddings are also less frequently accessed, hence the impact would be minimal. Our experiments confirm the intuition by showing 2.5\% increase in the total requests, while allowing for up to 2x cache size in some configurations in practice. We see up to 48\% increase in performance for cases where performance is bounded by user embeddings in \sm.



\begin{algorithm}
   \caption{de-pruning at load time}
   \label{alg:deprune_code}
\begin{algorithmic}
\small
   \STATE {\bfseries Input:} Tables
   \FOR{t in Tables}
   \IF{t is pruned-table}
   \STATE nt = new Table(dim=[t.mapper.dim[0], t.dim[1])
   \FOR{i in t.mapper}
   \IF{i is pruned-row}
   \STATE nt[i] = createZero(i.dim)
   \ELSE
   \STATE nt[i] = t[i]
   \ENDIF
   \ENDFOR
   \STATE t = nt
   \ENDIF
   \STATE SaveToSM(t)
   \ENDFOR

\end{algorithmic}
\end{algorithm}

\subsection{Placement}

With a software defined cache in \fm, there will be two choice to use the \fm space; 1) use all the available space for the cache 2) use portion of \fm to map tables directly, and portions for the cache.

\begin{table}[]
\caption{\small{Placement Strategies.}}
\vspace{2mm}
\label{tab:placement}
\tiny
\begin{tabular}{l|l}
\toprule
Policy & Description \\ \midrule
\sm only with Cache & all the (user) tables are mapped to \sm. \\ 
& rely on Cache in \fm to keep the hot rows in faster memory \\ \midrule
Fixed \fm, \sm with Cache & some tables could be directly mapped \\
&to \fm based on a given policy. The rest will be placed on \sm\\ \midrule
per table cache enablement & For \sm cache in \fm. Low temporal locality \\
&tables will not use the cache. \\ \midrule
\end{tabular}
\vspace{-3mm}
\end{table}

In general, allocating all the tables to \sm and relying on the cache to keep the hot rows in \fm will perform well across the board. However, given the extra overhead of looking up an embedding row from the cache vs plane memory, there are possibilities to improve the performance further with more detailed placement. Table~\ref{tab:placement} lists different placement categories. Figure~\ref{fig:cache_org} shows the impact of placement with different budget for direct placement on DRAM on a 150GB model running inferenceEval (which is more sensitive to placement than inference because the user and item batch sizes are the same). 



\textit{Tuning API}: Pre-defined placement policies based on table size and pooling factor can be enabled. We also implemented an option to providing a list of tables which should not be placed in \sm (for more elaborate offline placement). All placement policies adhere to a configurable DRAM budget to place tables on DRAM directly.

\subsection{Experiment Setup}
We consider 3 models with different characteristics which reflects models in use for different usecase.

We use a set of hardware platforms to evaluate different models, as listed in Table~\ref{tab:hw}. The choice of platform is driven by the usecase characteristics and requirements. Hence, some of the possible evaluation combinations are not feasible (e.g running the exact HW with and without using SSD).

\begin{table}[]
\small
\caption{\small{Target models configuration}}
\vspace{2mm}
\label{tab:models}
\footnotesize
\begin{tabular}{l|ccc}
\toprule
Model & \igreel & \ctrinsta & \adsother   \\ \toprule
Num parameters & 143B & 450B & 5T   \\ 
Size (GB) & 143 & 150 & 1000 \\ \midrule
Num of user emb tables & 61 & 450 & 1800  \\ 
Emb table dim (B) & [90, 172] & [32, 288] & [32, 512] \\ 
\tiny{(range [min, max], avg)} & avg: 51 & avg: 64 & avg: 192  \\ 
Avg pooling factor (PF) & 42 & 25 & 26   \\ 
User batch & 1 & 1 & 1 \\ \midrule
Num of item emb tables & 30 & 280 & 900  \\ 
Emb table dim (B) & [90, 172] & [4, 320] & [32, 512] \\ 
\tiny{(range [min, max], avg)} & avg: 69 & avg: 38 & avg: 192  \\ 
Avg pooling factor (PF) & 9 & 14 & 26   \\ 
Item batch & 50 & 150 & 1000 \\ \midrule
Num MLP layers & 31 & 43 & 35  \\ 
Avg MLP size & 300 & 735 & 6000  \\
\bottomrule
\end{tabular}
\end{table}

\begin{table}[!t]
\small
\caption{\small{Hw configurations. All the CPUs are Intel. For SSDs, N stands for Nand Flash, and O for Optane SSDs. See~\cite{fbaccelerator} for more information on the accelerators.}}
\vspace{2mm}
\label{tab:hw}
\centering
\setlength{\tabcolsep}{1.6mm}{
\begin{tabular}{l|c|c|c|c}
\toprule
Name & CPU & DRAM (GB) & SSD & Accelerator \\ \midrule
\Tsix & 2xXeon & 256 & - & - \\ \midrule
\Tone & 1xXeon & 64 & - & - \\ \midrule
\Tthree & 1xXeon & 64& 2x2TB N & - \\ \midrule
\Tseventeen & 1xXeon & 64& 2x1TB N & Yes \\ \midrule 
\Tseventeenoptane & 1xXeon & 64& 2x0.4TB O & Yes \\
\bottomrule
\end{tabular}}
\vspace{-3mm}
\end{table}

\section{Results}

\subsection{Using simpler HW}
In many occasions, a usecase has to select from a very limited set of available host types deployed in DC. Such different host types provide different CPU, DRAM, and Storage capabilities. Using \sm for \igreel allows for lowering the DRAM capacity requirement per host to serve a model. This enables using single socket, 64GB DRAM \Tthree instead of dual socket 256GB DRAM \Tsix. While each \Tthree can sustain lower QPS at the desired latency compared to \Tsix, the more favorable compute to DRAM ratio of \Tthree plus having attached SSDs leads to 20\% lower power consumption considering the full scale of the serving. Table~\ref{tab:ig_t3} shows the results. 

The IOPS required by the model at 120 QPS is around 246K ($120 QPS \times 50 Tables \times 42 avg PF$). We observe cache hit rate of more than 96\% in steady state which typically is reached within a few minutes after a full model update. This means less than 10K IOPS in steady state. 

We observe higher p99 latency on \Tthree due to occasional long tail latency of Nand Flash. Nonetheless p95 is the metric of interest for this usecase, which is matched on \Tthree.  Using \Tthree saves equivalent of 159.4 TB of DRAM for this particular model in production like settings.

\begin{table}
\caption{\small{Impact of using SSD on saving power consumption. Reported power is normalized. Deployment with SSD can reach the same latency as deployment on DRAM only, resulting on 20\% power saving.}}
\vspace{2mm}
\small
\label{tab:ig_t3}
\begin{tabular}{l|c|c|c|c}
\toprule
Scenario & QPS & Power &  Total Hosts & Total Power \\ \midrule
\Tsix & 240 & 1.0 & 1200 & 1200 \\ \midrule
\Tthree + SDM & 120 & 0.4 & 2400 & 960 \\ \bottomrule
\end{tabular}
\vspace{-3mm}
\end{table}

\subsection{Avoiding Scale-Out}
\ctrinsta uses an accelerator enabled platform (\Tseventeen) due to its higher compute intensity~\cite{anderson2021first}. The item embeddings as well as the dense part of the model is mapped to the accelerator. The user embeddings are mapped to the host CPU. \Tseventeen has adequate accelerator memory to host the item embeddings, however, the 64GB host DRAM is smaller than the 100GB memory required by the user embeddings. The extra memory required for the user embeddings is achieved through the scale out as presented in~\cite{lui2021understanding}, using \Tone host types. A \Tone on average can serve 5 \Tseventeen. 

For this usecase, using \sm prevents scale out. However, given the accelerated QPS per host, a higher degree of IOPS is required from \sm ($450 QPS \times 450 Tables \times 25 Avg PF = 4.8 MIOPS$). We observe more than 90\% hit rate in the \sm cache. So the average sustained IOPS required is around 480 kIOPS. As shown in Table~\ref{tab:ads_optane}, the two Nand flash on \Tseventeen provide aggregate minimal IOPS of around 1M. However, due to long latency accessing nand flash, we have to considerably underutilize the devices to keep the latency low. Hence in practice Nand Flash in this setup considerably impacts QPS. However, Optane SSD provide much higher IOPS and lower latency, keeping the user embedding processing out of the critical path. By removing the need to scale out, \Tseventeenoptane reduces the power consumption by 5\%. 

At the same time, \Tseventeenoptane simplifies the serving paradigm, as the scale out paradigm is more complex to operate, and more prone to failures given that many more hosts are involved in serving a single query. While the power saving is modest, it increases as the models grow. 


\begin{table}
\caption{\small{Impact of using SSD on power consumption for \ctrinsta}}
\label{tab:ads_optane}
\vspace{2mm}
\footnotesize
\resizebox{\columnwidth}{!}{
\begin{tabular}{l|c|c|c|c}
\toprule
Scenario & QPS & Power & Total Hosts & Total Power \\ \midrule
\Tseventeen + ScaleOut & 450 & 1.0 + 0.25 & 1500 + 300 & 1575  \\ \midrule
\Tseventeen + SDM & 230 & 1.4 & 2978 & 2978 \\ \midrule
\Tseventeenoptane + SDM & 450 & 1.0 & 1500 & 1500 \\ \bottomrule
\end{tabular}}
\vspace{-3mm}
\end{table}

\subsection{Facilitate Multi-Tenancy}
For \adsother we present the estimated results, as it is a future use case, with the chance to impact the design of the host type. \adsother represents a future model which could run on an updated accelerator-enabled platform(e.g. see ~\cite{zion}). The primary arguments for SDM in such platform is to limit the amount of DRAM deployed per host. This is primarily a cost argument as DRAM power is not a significant portion of the total power in such platforms. The power savings come from allowing for increased accelerator utilization without becoming DRAM memory capacity bound through Multi-tenancy.

Multi-tenancy refers to running multiple models on the same host. This capability is becoming more important (e.g. see~\cite{jouppi2021ten}) as it allows for co-locating models with different requirements, and balancing the utilization of different resources such as accelerator, CPU, and DRAM. The balanced utilization leads to increase overall host utilization, and power saving. As an example, at any given time, there are a large number of experimental models running, of which, a subset will eventually be promoted for full scale deployment. Our observation is that given the number of experimental models running per production models, and on average it consumes up to quarter of the allocated resourced. Such experimental models run on a small volume of traffic, and hence have low QPS requirement per model, which could leave the hosts underutilized. This becomes more important as more compute capability is packed into a single host with the advent of more powerful accelerators, increasing the cost of a model underutilizing a host. Co-locating more than one model on a given host increases utilization. Notably, the memory capacity requirement will scale with the number of models co-located together. Therefore, serving becomes memory capacity bound.

Using \sm in this case prevents the memory capacity boundness due to the multi-tenancy, by increasing memory capacity available to the models per host.

To drive the \sm capacity and BW requirement per host, we use \adsother as the representative model. We estimate the QPS on the target hardware by measuring the QPS on an available similar hardware, and extrapolating the QPS based on the expected increase in compute and BW. The BW needed from \sm could be calculated according to Equation~\ref{eq:bw_ro}. Table~\ref{tab:ads_other_hw} shows the need for 36 MIOPS which could be satisfied by 9 OptaneSSD, each providing 4 MIOPS. 

Given the number of experimental models and their required QPS, we observe 63\% utilization of the hosts at the scale. Table~\ref{tab:ads_other} shows the roofline estimation for power saving with multi-tenancy enabled through leveraging slower memory. The modeling shows up to 29\% power saving.

\begin{table}
\caption{\small{SDM-based HW configuration for future models (\adsother). Number of SSDs are derived based on BW, and hence IOPS needed for the user embedding tables (IOPS bound).}}
\vspace{2mm}
\tiny
\label{tab:ads_other_hw}
\setlength{\tabcolsep}{1.9mm}{
\begin{tabular}{l|c|c|c|c|c|c|c}
\toprule
Model & QPS &  User Tables & PF & Emb dim & Hit Rate & MIOPS & numSSDs \\ \midrule
\adsother & 3150 & 2000 & 30 & 512 & 80\% & 36 & 9 \\ \midrule
\end{tabular}}
\vspace{-8mm}
\end{table}

\begin{table}
\caption{\small{Using SDM with \sm allows for placing multiple experimental models on the same host, hence increasing aggregate utilization of the fleet. This results in better perf/watt, and power saving for the given model class. The increase in the power consumption of the host is based on the OptaneSSDs needed.}}
\vspace{2mm}
\centering
\footnotesize
\label{tab:ads_other}
\setlength{\tabcolsep}{3mm}{
\begin{tabular}{l|c|c|c}
\toprule
Scenario & Power & Utilization & fleet power \\ \midrule
\Tsixteen & 1.0 & 0.63 & 1.0  \\ \midrule
\Tsixteenoptane + SDM & 1.01 & 0.90 & 0.71 \\ \midrule
\end{tabular}}
\vspace{-3mm}
\end{table}

\section {Related work}
SSDs have been used to extend memory in different applications. For example~\cite{heo2021boss} use SSD to increase memory capacity for search applications. ~\cite{zhao2020distributed} develops a distributed training system using SSD in a hierarchical fashion to increase available memory capacity. Inference, however, is more latency sensitive compared to training, which makes it harder to leverage SSD. 

 \cite{eisenman2018bandana} is among the pioneers tackling the challenges in using SSD for inference. It groups the embedding vectors of a given tables to increase the possibility that the grouped embeddings could be accessed together. This helps reduce the read amplification due to large device block size read, which is considerably bigger than embedding dimension. Our work does not follow this path due to the implication of grouping on the latency between model updates. In~\cite{wilkening2021recssd} authors leverage the limited compute and DRAM in the SSD controller to collect and pack requested embeddings across different pages, hence making the data transfer over PCIe more efficient. In our work, we pursue techniques to reduce the access granularity which addresses read amplification and inefficient use of the BW by reading large block size. ~\cite{liu2021jizhi} present a recommendation system which can leverage SSD for embeddings. While they mention the implication of using SSD on latency, they do not further discuss how to remedy such increase in latency, or exact implication of using SSD vs memory. \cite{lui2021understanding} use scale out and shard the model across multiple servers to scale the memory capacity available to the usecase.
 
\section{Conclusion}
Rapid increase in Deep Learning Recommendation Model (DLRM) size makes it more expensive in terms of cost and power to serve such models. Power, is particularly among the most important metrics at DC scale. Companies operating DCs are willing to pay for extra compute, but the rate of growth is limited by the rate at which the power could be provisioned. We leverage the inherit skew in BW among different embedding tables in DLRM to deploy a Software Defined tiered memory increasing memory capacity per host by leveraging Storage Class Memory. We evaluate and address a range of challenges, such as fast IO, capturing locality, trade-off of capacity vs compute and BW. We discuss the value of such technology under different deployment scenarios. We observe 20\% power saving serving a large model while using a \textit{simpler hardware} with Nand Flash, 5\% power saving using another compute heavy model by \textit{avoiding scale-out}, and projected 29\% improvement in perf/watt by increasing utilization of Accelerator enabled platforms through \textit{multi-tenancy} using Optane SSD. Such power and perf/watt optimizations are considerable given the power boundness of serving such models at DC scale.  

\section*{Acknowledgements}
We would like to acknowledge the valuable insight and help from \emph{Assaf Eisenman, Jeremy Yang, Jesseh Koh, Mo Cao, Alex Kachurin, Pavan Shetty, Kiran Malwankar, Yinghai Lu, Pallab Bhattacharya, Lu Fang, and Ajay Somani}.

\bibliography{main}
\bibliographystyle{mlsys2022}

\appendix
\section{Discussion}
\label{sec:discussion}

\subsection{CPU cost of high IOPS}

At very high IO rate, there is always data returned in the completion queues to be processed. This is particularly the case with Optane SSD, given high IOPS offered per device~\footnote{Number of outstanding IOs when using OptaneSSD =  $inflyQueries*numTables*avgPoolingFactor$}. Hence removing the IRQ overhead and performing polling based IO at the OS side could show better performance for both latency and IOPS/Core. We observe 50\% improvement on IOPS/Core when enabling polling. However, we found it prohibitively complex to enable polling for our usecase. This is because operator based execution in Caffe2 or Pytorch would not allow for creating producer-consumer pool across all the embedding operators in the model.

Newer technologies to expose SSD space to the CPU might alleviate this problem. Ideally standardization of such solution (e.g. through CXL) would make the option more adoptable.

\subsection{Inter-Op Parallelism}
The SparseLengthSum operator in Caffe2 or EmbeddingBag operator in Pytorch could involve IO when the embedding tables associated with the operator are placed on \sm. Hence it becomes important to not only enable async IO for access to embeddings for a given table, but also provide async execution of the operators in such Deep Learning platforms. Such inter-Op parallelism allows for more efficient discovery of IOs that need to be issued, and enable IO and computation overlap. Therefore, the inter-Op parallelism reduces latency per query. In a latency sensitive usecase, higher latency per query could result in underutilization of the host, and hence lower throughput. Therefor, inter-Op parallelism also improves throughput. For example we have observed 20\% reduction in latency per query through inter-Op parallelism, resulting in 20\% more QPS per host at the desired latency for model \igreel (Table~\ref{tab:models}).



\subsection{Model update}

Models are refreshed frequently, with a desire for more frequent updates (e.g updates every few minutes), to keep the models as up to date as possible. However, given the large model size, updates could be separated to updating dense parameters and updating embedding tables, which could happen with different frequency (embedding updates being less frequent). Incremental update is another path to increasing update frequency for the model. Given the need to save the embeddings into \sm, the time it takes to update the model will be increased. Hence, incremental updates are considered to minimize the amount of data that needs to be updated.

As new weights  stream in, the host could be offline or still online serving traffic. The former prevent mixture of read and write BW which would considerably impact performance of Nand flash. The latter would allow for better utilization of the resources. Given the software defined cache, we can update the cache first and allow for dirty write backs to update the \sm. 

Section~\ref{sec:scm_choice} discusses how endurance could limit model update frequency.

\subsection{Warmup}

Cold \sm cache in \fm could impact the performance right after a full model update. We observe that caches warmup in order of a few minutes. But the perf impact need to be compensated by over-provisioning the capacity. For example if 1) r=10\% of the hosts serving a model are being updated in a given time (rolling update), and 2) the performance during warmup is p=50\% of steady state, 3) update every t=30 minutes, 4) warmup in w=5 minutes, we need $(r * w) / (p * t) = (10\% * 30) / (50\% * 5) = 1.2\%$ more capacity to offset the slowdown.

\subsection{de-quantization}
Quantization is a widely use technique~\cite{han2015deep}, and for embedding tables it helps to reduce model size as well as memory BW. Given the higher \sm capacity, we can dequantize the embedding table at loading time into the \sm, and save the dequantization at run time. It will consume more memory space in \sm, which typically is not memory capacity bound. It will not add to BW consumption of \sm in some of the systems due to higher access granularity to \sm (e.g. 72B int8 qunatized embedding with 64 embedding elements and 8 byte quantization parameters per row expand to 256B, still smaller than access granularity for Nand Flash). However, dequantization at loading leads to less efficient use of \fm space for cache. This is because less number of embedding rows could be stored in a given cache size when each row enlarges due to dequantization. We observe that while under very CPU bound usecases dequantization could help, but for most of the usecases the impact on cache is dominant and does not lead to benefit. Pooled embedding cache (Section ~\ref{sec:pooledemb}) provides a more fine tuned solution which can leverage dequantized (and pooled) embeddings in a more selective manner.

%


\end{document}